\documentclass[12pt] {article}
\usepackage{latexsym}{\rm }
\usepackage{graphics}
\usepackage{graphpap}

\newcommand{\beq}{\begin{equation}}
\newcommand{\feq}[1]{\label{#1} \end{equation}}
\newcommand{\beqr}{\begin{eqnarray}}
\newcommand{\feqr}{\end{eqnarray}}
\def\non{\nonumber}
\def\noi{\noindent}

\newcommand{\rf}[1]{(\ref{#1})}

\def\np#1#2#3{Nucl. Phys. {\bf{B#1}} (#2) #3}

\def\plb#1#2#3{Phys. Lett. {\bf{B#1}} (#2) #3}

\renewcommand{\thefootnote}{\fnsymbol{footnote}}

\begin{document}

\begin{center}


{\Large \bf Instantons in Four-Fermi Term Broken SUSY with General Potential. }\\
[4mm]

\large{Agapitos Hatzinikitas} \\ [5mm]

{\small University of Crete, \\
Department of Applied Mathematics, \\
L. Knosou-Ambelokipi, 71409 Iraklio Crete,\\
Greece, \\
Email: ahatzini@tem.uoc.gr}\\ [5mm]

\large{and} \\ [5mm]

\large{Ioannis Smyrnakis} \\ [5mm]

{\small University of Crete, \\
Department of Applied Mathematics, \\
L. Knosou-Ambelokipi, 71409 Iraklio Crete,\\
Greece, \\
Email: smyrnaki@tem.uoc.gr} \vspace{5mm}

\end{center}

\begin{abstract}
It is shown how to solve the Euclidean equations of motion of a point particle in a general potential and in the presence 
of a four-Fermi term. The classical action in this theory depends explicitly on a set of four fermionic collective 
coordinates. The corrections to the classical action due to the presence of fermions are of topological nature in 
the sense that they depend only on the values of the fields at the boundary points $\tau \rightarrow \pm \infty$. As an 
application, the Sine-Gordon model with a four-Fermi term is solved explicitly and the corrections to the classical action 
are computed.    
\end{abstract}
\newpage

\section{Introduction}
It is well-known that the equations of motion for a point particle in Euclidean space moving under the influence of a 
Minkowski potential possessing at least two degenerate minima, admit finite action solutions which are the instantons. 
Their existence
is responsible for tunnelling processes in the Minkowski space. The symmetries of the action are reflected in the space of 
instanton solutions.
Local deformations of these solutions in the directions determined by the symmetries give rise to zero modes in the 
semiclassical expansion. To avoid infinities in the path integral due to these zero modes one is forced to introduce 
collective coordinates \cite{rajaraman, shifman}. 
\par In the presence of rigid supersymmetry the above solutions are still instantons provided that the fermions vanish 
\cite{ferrara,witten}. Applying the rigid 
supersymmetry transformation rules on these instantons one determines a more complete set of instantons where the fermions 
no longer vanish 
but instead they depend linearly on Grassmann collective coordinates. The number of Grassmann collective coordinates is equal to
the number of the Fermi fields present. Since the new instantons are related to the previous ones by supersymmetry, the 
classical action remains the same. 
\par It is possible to break supersymmetry through the introduction of a four-Fermi term \cite{peter}. 
The new equations of motion
can be solved iteratively starting with the instanton solution in the supersymmetric case. The iterative process terminates
due to the nature of Grassmann collective coordinates so in this way it is possible to obtain exact solutions to the new 
equations of motion. Since there is no symmetry involved
in obtaining these solutions the instanton action will change. The correction is the integral of a total derivative term,
so it depends only on the boundary values of the fields. It depends also on the Grassmann collective coordinates introduced 
by supersymmetry. Note that the fermionic fields become infinite as $\tau \rightarrow \pm \infty$ while it is possible 
to keep the bosonic field finite by appropriate choice of the integration constants. Nevertheless, despite this infinity,
the action remains finite. 
\par As an explicit example we consider the quantum mechanical Sine-Gordon potential with a four-Fermi term. The iterative
solution of the equations of motion is demonstrated explicitly determining in this way the instantons. The integration 
constant that renders the bosonic field finite is determined. In this case the instanton action
becomes $S=-\frac{8m^3}{\lambda}+\epsilon^{ijkl}\xi_i \xi_j \xi_k \xi_l \frac{mg}{12}$ where $g, \lambda$ are coupling 
constants.
\par Finally, we solve the equations of motion for a general potential in terms of 
the bosonic part of the instanton when the Fermi field is zero, which is used as a new variable instead of $\tau$. 
It is interesting that in order to 
compute the action corrections it is not necessary to solve the nonlinear BPS equation.
 In fact it is possible to express completely both the instanton and the finite boson integration 
constant in terms of the new variable.

\renewcommand{\thefootnote}{\arabic{footnote}}
\setcounter{footnote}{0}

\section{The Quantum Mechanical Model}

We start by considering the following one dimensional quantum mechanical model in Euclidean space:
\beqr
\label{action}
S_{cl}= -\frac{1}{2}\int_{-\infty}^{\infty}\Bigl[ (\dot{x}(\tau))^2 + U^2(x)\Bigr]d\tau. 
\feqr
 The potential $-\frac{1}{2}U^2(x)$ of the equivalent particle is asumed to have a 
number of degenerate minima. 
\par The equation of motion is
\beqr
\label{eqb1}
\ddot{x}-U(x)U^{\prime}(x)=0.
\feqr  
The instanton solution of this equation satisfies the BPS equation
\beqr
\label{bps}
\dot{x}_{in}+U(x_{in})=0
\feqr
subject to the conditions $x_{in}(\pm \infty)= C_{\pm}$ where $U(C_{\pm})=0$.
The action is invariant under time translations which implies that if $x_{in}(\tau)$ is a solution to the BPS equation then 
$x_{in}(\tau-\tau_0)$ is also a solution. This means that
\beqr
\label{zerom} 
\tilde{Z}_0(\tau-\tau_0)=\dot{x}_{in}(\tau - \tau_0)=- \frac{d}{d\tau_0}x_{in}(\tau -\tau_0)
\feqr
is a zero mode of the operator corresponding to the quadratic variation of the action around $x_{in}$. It
satisfies the equation 
\beqr
\label{zerob}
\dot{\tilde{Z}}_0(\tau-\tau_0) + U^{\prime}(x_{in})\tilde{Z}_0(\tau - \tau_0)= 0. 
\feqr
Note that the normalization of $\tilde{Z}_0$ defined as $\dot{x}_{in}$ is just the absolute value of the classical action. 
This is so because
\beqr
\label{bfin}
S_{cl}=-\int_{-\infty}^{\infty}(\dot{x}_{in})^2 d\tau=-\int_{-\infty}^{\infty}\tilde{Z}_0^2 d\tau
\feqr
where we use the BPS equation. So it is reasonable to define the normalised zero mode as follows
\beqr
\label{newz0}
Z_0=|S_{cl}|^{-1/2}\tilde{Z}_0. 
\feqr
\par It is possible to introduce fermions into this model by adding the terms
\beqr
\label{2f}
S_{2f}=-\frac{1}{2}\int_{-\infty}^{\infty}\Bigl[ \psi_{i}^{T}\dot{\psi}_i
+ (\psi_{i}^{T}\sigma_{2}\psi_i)U^{\prime} \Bigr]d\tau. 
\feqr 
where $\sigma_2=\left( \begin{array}{cl} 0 & -i \\ i & 0 \end{array} \right)$, $\psi_i$ are two component Majorana fermions 
and the fermionic index is a colour index 
ranging in $i=1,\cdots ,4$. 
\par Each fermion is related to a boson through rigid $N=1$ supersymmetry realised by the transformations
\beqr
\label{susytr}
\delta x = \epsilon^{T}\sigma_{2}\psi; \quad \delta \psi = \sigma_2 \dot{x}\epsilon -U \epsilon.
\feqr  
\noi The spinor $\epsilon$ can be expanded in terms of the eigenstates 
$\psi_\pm=\frac{1}{\sqrt{2}}\left(\begin{array}{c} 1 \\ \pm i\end{array}\right)$ of 
$\sigma_2$ as follows
\beqr
\label{spex}
\epsilon=\frac{1}{2} \left( \xi_+ \psi_+ + \xi_- \psi_- \right)
\feqr 
and then it can be proved that if $\epsilon=(\frac{1+\sigma_2}{2})\epsilon$ then 
\beqr
\label{susyz}
\delta x=0; \quad \delta \psi=\xi_+ \psi_+ Z_0 (\tau).
\feqr
Starting from the configuration $x=x_{in}$, $\psi =0$ and integrating over the above supersymmetry transformations 
we arrive at the instanton given by $x=x_{in}$ and
\beqr
\label{zeroi}
\psi_i^{(1)}=\xi_{i} Z_0(\tau-\tau_0)\psi_+.
\feqr 
To each colour index we associate the same bosonic zero mode but different Grasmannian collective coordinates $\xi_i$.
Following \cite{peter} we add to the action a four Fermi term which breaks supersymmetry 
\beqr
\label{4f}
S_{4f}=\frac{g}{4} \int_{-\infty}^{\infty} \epsilon_{ijkl}(\psi_{i}^{T}\sigma_{1}\psi_j)(\psi_{k}^{T}\sigma_{1}\psi_l)d\tau
\feqr
where $\sigma_1=\left( \begin{array}{cl} 0 & 1 \\ 1 & 0 \end{array} \right)$. The new field equations are now
\beqr
\label{eqbfn}
\ddot{x} - U U^{\prime}&=& \frac{1}{2}(\psi_i^T \sigma_2 \psi_i)U^{\prime \prime} \\
\dot{\psi}_i + \sigma_2 \psi_i U^{\prime} &=& g\epsilon_{ijkl}\sigma_1 \psi_j (\psi_k^T \sigma_1 \psi_l).
\feqr
Expanding $\psi_i$ w.r.t the GCC one has $\psi_i=\psi_i^{(1)}+\psi_i^{(3)}$. Note that all the other terms in 
the expansion vanish. Making the ansatz
\beqr
\label{ansatz}
\psi_i^{(3)}=\alpha(\tau)\epsilon_{ijkl}\xi_j \xi_k \xi_l \psi_-,
\feqr 
and plugging it into the fermionic field equation (15) one gets
\beqr
\label{ferm3}
\dot{\alpha}-\alpha U^{\prime}=-g Z_0^3.
\feqr
Since the solution of the homogeneous equation is $Z_0^{-1}$ it is natural to set 
\beqr
\label{soluy}
\alpha(\tau)=Z_0^{-1}y(\tau).
\feqr
This leads to the equation
\beqr
\label{eqy}
\dot{y}(\tau)=-g Z_0^4.
\feqr
Similarly we can expand $x(\tau)=x_{in}(\tau)+x^{(4)}(\tau)$ and plug this into the bosonic field equation which gives
\beqr
\label{bos4}
\ddot{x}^{(4)}-x^{(4)}(UU^{\prime \prime}+U^{\prime 2})(x_{in})=-\epsilon \xi^4 Z_0 \alpha(\tau)U^{\prime \prime}(x_{in}).
\feqr
The solution of the homogeneous equation is $Z_0(\tau)$ so it is natural to set 
$x^{(4)}= \epsilon \xi^4 Z_0(\tau) \beta(\tau)$ where $\beta(\tau)$ satisfies
\beqr
\label{betaeq}
\frac{d}{d\tau}(Z_0^2 \dot{\beta})= -y(\tau)Z_0(\tau)U^{\prime \prime}(x_{in}) 
\feqr
and $\epsilon \xi^4=\epsilon_{ijkl}\xi_i \xi_j \xi_k \xi_l$. 
\par It is interesting to compute the corrections to the classical action due to $\psi_i^{(3)}$ and $x^{(4)}$. The two
Fermi term gives
\beqr
S_{2f} &=& -\frac{1}{2}\int_{-\infty}^{\infty} \Bigl[ \psi_i^{T(1)}\dot{\psi}_i^{(3)} +\psi_i^{T(3)}\dot{\psi}_i^{(1)} 
+ (\psi_i^{T(1)}\sigma_2 \psi_i^{(3)}+\psi_i^{T(3)}\sigma_2 \psi_i^{(1)})
U^{\prime}(x_{in}) \Bigr]\non \\
&=& -\frac{1}{2}\int_{-\infty}^{\infty} \Bigl[\psi_i^{T(3)}(\dot{\psi}_i^{(1)}+\sigma_2 \psi_i^{(1)}U^{\prime})+
\psi_i^{T(1)}(\dot{\psi}_i^{(3)}+\sigma_2 \psi_i^{(3)}U^{\prime})\Bigr] \non \\
&=& -\frac{1}{2}\left( \epsilon \xi^4 \right) \int_{-\infty}^{\infty} \dot{y} d\tau 
=-\frac{1}{2}\left( \epsilon \xi^4 \right) \left(y(\infty)-y(-\infty) \right)
\label{2ffin}
\feqr 
and the four Fermi term 
\beqr
\label{4ffn}
S_{4f}=\frac{g}{4}\int_{-\infty}^{\infty}\epsilon_{ijkl} (\psi_i^{T(1)}\sigma_1 \psi_j^{(1)}) 
(\psi_k^{T(1)}\sigma_1 \psi_l^{(1)})=\frac{1}{4}\left( \epsilon \xi^4 \right) \left(y(\infty)-y(-\infty) \right). 
\feqr 
One easily checks that $S_{4f}=-\frac{1}{2}S_{2f}$. 
\par The bosonic correction gives
\beqr
\label{bosc}
S_{bc}&=& -\int_{-\infty}^{\infty}\Bigl[\dot{x}_{in}\dot{x}^{(4)}+ x^{(4)}U(x_{in})U^{\prime}(x_{in}) \Bigr] \non \\
&=& -\int_{-\infty}^{\infty}\frac{d}{d\tau}(x^{(4)}\dot{x}_{in})d\tau
=-\epsilon \xi^4 \sqrt{S_{cl}} \left(\lim_{\tau\rightarrow \infty}(Z_0^2(\tau)\beta(\tau))
- \lim_{\tau \rightarrow -\infty}(Z_0^2(\tau)\beta(\tau))\right). 
\feqr
It is worth noting that if the bosonic field is finite as $ \tau=\pm\infty $ then 
$ \lim_{\tau\rightarrow \infty} \left( \beta (\tau) Z_0 (\tau) \right) $ is finite and since 
$ \lim_{\tau \rightarrow \infty}Z_0 (\tau)=0 $
the bosonic correction vanishes.

\section{The Sine-Gordon model}

For the Sine-Gordon model the action is
\beqr
\label{sg1}
S^{SG}_{cl}=-\frac{1}{2}\int_{-\infty}^{\infty}\Bigl[\dot{x}^2 + \frac{2m^4}{\lambda} 
(1-\cos(\frac{\sqrt{\lambda}}{m}x)) \Bigr] d\tau. 
\feqr
Here 
\beqr
\label{uid}
U(x)=\frac{2m^2}{\sqrt{\lambda}}\sin(\frac{\sqrt{\lambda}}{2m}x).
\feqr
The BPS equation takes the form 
\beqr
\label{bps2}
\dot{x}_{in}+\frac{2m^2}{\sqrt{\lambda}}\sin{\frac{\sqrt{\lambda}}{2m}x_{in}}=0
\feqr
and can be easily solved to give 
\beqr 
\label{sginst}
x_{in}(\tau )=\pm 4\frac{m}{\sqrt{\lambda}}\tan^{-1}e^{-m(\tau-\tau_0)}.
\feqr
The minus sign corresponds to the instanton solution and the plus sign to the anti-instanton.  
In what follows we are going to work with the instanton.  
The zero mode corresponding to the instanton is 
\beqr
\label{zero2}
\tilde{Z}_0(\tau )=\frac{dx_{in}}{d\tau}=2\frac{m^2}{\sqrt{\lambda}}\frac{1}{cosh(m(\tau-\tau_0))}
\feqr 
and the classical action becomes 
\beqr
\label{clact2}
S_{cl}=-\int_{-\infty }^{\infty }\tilde{Z}_0^2d\tau =-\frac{8m^3}{\lambda}
\feqr 
So 
\beqr
\label{zsg}
Z_0=\sqrt{\frac{m}{2}}\frac{1}{cosh(m(\tau - \tau_0))}.
\feqr
Upon introducing the two and four Fermi terms given in the previous section, we get that the fermionic field is 
$\psi_i=\psi_i^{(1)}+\psi_i^{(3)}$ where $\psi_i^{(1)}$ is given by \rf{zeroi} and $\psi_i^{(3)}$ is given by \rf{ansatz}.
The function $y(\tau )$ is the solution of the equation 
\beqr 
\label{y2}
\dot{y}(\tau)=-gZ_0^4(\tau)=-g\frac{m^2}{4}\frac{1}{cosh^4(m(\tau-\tau_0))}
\feqr 
This can be solved by setting $z=tanh(m(\tau -\tau_0))$.  With this change of variable the solution is written as
\beqr
\label{y3}
y(z)=-\frac{gm}{4}(a+z-\frac{1}{3}z^3)
\feqr 
and thus 
\beqr
\label{psi1}
\psi^{(3)}_i=-\frac{g}{2}\sqrt{\frac{m}{2}}\frac{1}{\sqrt{1-z^2}}(a+z-\frac{1}{3}z^3)\epsilon_{ijkl}\xi_j\xi_k\xi_l
\psi_-.
\feqr
To determine the form of $x^{(4)}$ we solve the bosonic field equation \rf{betaeq}. In terms of the new variable z
this equation becomes
\beqr
\label{eqz}
\frac{d}{dz}\left(\frac{d\beta}{dz} Z_0^4(\tau) \right)=\frac{1}{4}\sqrt{\frac{\lambda}{2m}}y(\tau).
\feqr
This gives 
\beqr
\label{dbdz}
\frac{d\beta}{dz}=-\frac{1}{4}\frac{g}{m}\sqrt{\frac{\lambda}{2m}} \left(\frac{A}{(1-z^2)^2} 
+ \frac{\alpha z}{(1-z^2)^2}+ \frac{z^2}{2(1-z^2)^2} - \frac{z^4}{12(1-z^2)^2} \right).
\feqr
The solution to this equation is
\beqr
\label{solx4}
\beta(z)= \frac{1}{4} \frac{g}{m} \sqrt{\frac{\lambda}{2m}} \Bigl[ \left(\frac{A}{2}+\frac{5}{24}\right) 
\frac{z}{z^2 -1} - \left(\frac{A}{4}-\frac{1}{16} \right) \ln \left(\frac{z+1}{1-z} \right) + \frac{1}{12}z
+ \frac{1}{2} \frac{\alpha}{z^2 -1} +B \Bigr] 
\feqr
The contribution to the two and four Fermi terms is
\beqr
\label{24f}
S_{2f}+S_{4f}=\frac{1}{2}S_{2f}=\epsilon \xi^4 \frac{gm}{12}.
\feqr
The bosonic correction is
\beqr
\label{bosc3}
S_{bc}=-\epsilon \xi^4 \frac{1}{4}gm \left(A +\frac{5}{12} \right).
\feqr
If the bosonic field is bounded at infinity then $S_{bc}$ vanishes and this implies that $A=-\frac{5}{12}$. In this
case the full action is
\beqr
\label{fuact}
S_{tot}=-\frac{8m^3}{\lambda}+\epsilon \xi^4 \frac{gm}{12}.
\feqr

\section{Generalisation}

It is worth mentioning that it is possible to solve exactly the equations \rf{eqy}, \rf{betaeq} if we change variables
from $\tau$ to $x_{in}$. This is so because $\dot{x}_{in}=\sqrt{|S_{cl}|}Z_0$. Applying this change of variable to \rf{eqy}
we get
\beqr
\label{eqy1}  
\frac{dy}{dx_{in}}=-\frac{g}{\sqrt{|S_{cl}|}}Z_0^3=\frac{g}{|S_{cl}|^2}U^3(x_{in})
\feqr
where $|S_{cl}|=|\int_{C_-}^{C_+} U(x_{in})dx_{in}|$. This admits the solution 
\beqr
\label{soly1}
y=g m \tilde{\alpha} +\frac{g}{|S_{cl}|^2}\int U^3(x_{in}) dx_{in}.
\feqr
The integration constant $\tilde{\alpha}$ has been chosen to be dimensionless. Similarly equation \rf{betaeq} becomes
\beqr
\label{eqbta1} 
\frac{d}{dx_{in}}\left( \frac{d\beta}{dx_{in}} U^3(x_{in})\right)= \sqrt{|S_{cl}|} y(x_{in}) U^{\prime \prime }(x_{in}).
\feqr
Integrating once this equation and using \rf{eqy1} we get
\beqr
\label{sobet1}
\frac{d\beta}{dx_{in}}=\sqrt{|S_{cl}|} y(x_{in})\frac{U^{\prime}(x_{in})}{U^3(x_{in})}-\frac{g}{4} |S_{cl}|^{-\frac{3}{2}}
U(x_{in})+ gm^2 \sqrt{|S_{cl}|} \tilde{A} \frac{1}{U^3(x_{in})}.
\feqr
Integrating again we get
\beqr
\label{sob2}
\beta(x_{in})&=& -\frac{1}{2}|S_{cl}|^{-\frac{3}{2}}\frac{1}
{U^2(x_{in})} \left(gm |S_{cl}|^{2}\tilde{\alpha} +g \int U^3(x_{in}) dx_{in} \right) 
+\frac{1}{4} g |S_{cl}|^{-\frac{3}{2}}\int U(x_{in})dx_{in} \non \\
&+& g m^2 |S_{cl}|^{\frac{1}{2}}\tilde{A} \int \frac{1}{U^3(x_{in})} dx_{in} + g |S_{cl}|^{-\frac{1}{2}} \tilde{B}. 
\feqr
Recall now that as $\tau \rightarrow \pm\infty$, $x_{in} \rightarrow C_{\pm}$ where $U(C_{\pm})=0$. Demanding that 
the bosonic field correction is finite we get that $\lim_{x_{in}\rightarrow C_{\pm}}\left( \beta(x_{in})U(x_{in})\right)$ 
is finite,
so  $\lim_{x_{in}\rightarrow C_{\pm}}\left( \beta(x_{in})U^2(x_{in})\right)=0$. This translates into the conditions
\beqr
\label{con1}  
g m^2 |S_{cl}|^{\frac{1}{2}}\tilde{A} \lim_{x_{in}\rightarrow C_{+}}\left(U^2(x_{in})\int_{x_0}^{x_{in}} \frac{1}{U^3(s)}ds 
\right) -
\frac{g}{2}|S_{cl}|^{-\frac{3}{2}} \int_{x_0}^{C_+}U^3(s) ds =\frac{1}{2} \tilde{\alpha} g m |S_{cl}|^{\frac{1}{2}}\\
g m^2 |S_{cl}|^{\frac{1}{2}}\tilde{A} \lim_{x_{in}\rightarrow C_{-}}\left(U^2(x_{in})\int_{x_0}^{x_{in}} \frac{1}{U^3(s)}ds 
\right) -
\frac{g}{2}|S_{cl}|^{-\frac{3}{2}} \int_{x_0}^{C_-}U^3(s) ds =\frac{1}{2} \tilde{\alpha} g m |S_{cl}|^{\frac{1}{2}}  
\feqr
where the value of $\tilde{\alpha}$ is determined by the choice of $x_0$. Subtracting the two equations we arrive at
\beqr
\label{fin1}
\tilde{A}=\frac{1}{2 m^2}|S_{cl}|^{-2} \frac{\int_{C_-}^{C_+} U^3(s) ds}{\left(\lim_{x_{in}\rightarrow C_+} 
- \lim_{x_{in} \rightarrow
C_-} \right) \left(U^2(x_{in}) \int_{x_0}^{x_{in}}\frac{1}{U^3(s)}ds \right)}.
\feqr
These formulae have been checked in the cases of the double well and the Sine-Gordon potentials. For the double well 
potential the results agree with those of \cite{peter} provided that we redefine our integration constants appropriately. 
In the case of the Sine-Gordon model the results agree with those of the previous section under the following 
identification of the integration constants
\beqr
\label{idc} 
\tilde{A}=\frac{1}{16}-\frac{A}{4}; \quad \tilde{\alpha}=\frac{\alpha}{4}; \quad \tilde{B}=\frac{B}{2}.
\feqr  

\section{Conclusion} 
We have determined the corrections to the supersymmetric instanton, for a point particle in a general potential 
that admits at least two 
degenerate minima, due to the presence of a supesymmetry breaking four-Fermi term. 
Starting from the instanton solution of the 
supersymmetric case and applying an iterative procedure we obtain exact solution of the new equations of motion. There is 
no symmetry involved in obtaining these solutions so the classical action will receive corrections. 
If we demand that the bosonic field remains finite as $\tau \rightarrow \pm \infty$ then only the two- and four-fermion 
terms contribute 
corrections to the classical action. The fermionic fields diverge as $\tau \rightarrow \pm \infty$ nevertheless their 
corrections to the action remain finite. In the case of the Sine-Gordon potential  
the classical action gets modified by the contribution 
$\epsilon^{ijkl}\xi_i \xi_j \xi_k \xi_l \frac{mg}{12}$ where the $\xi_i$ are fermionic collective coordinates. 
Finally, by a suitable change of variables, 
we determine the corrections to the classical action for a general potential and we
compute the integration constant $A$ that makes the bosonic field finite
when $\tau \rightarrow \pm \infty$ in terms of the potential only.     

\bibliographystyle{plain}

\end{document}